\documentstyle[12pt,a4]{article}
\begin{document}
\def\doublespaced{\baselineskip=\normalbaselineskip\multiply\baselineskip
  by 150\divide\baselineskip by 100}
\doublespaced
\def\lsim{~{\rlap{\lower 3.5pt\hbox{$\mathchar\sim$}}\raise 1pt\hbox{$<$}}\,}
\def\gsim{~{\rlap{\lower 3.5pt\hbox{$\mathchar\sim$}}\raise 1pt\hbox{$>$}}\,}
\def\thisday{~\today ~and~ hep-ph/0003036~~}


\begin{titlepage}
\vspace{0.5cm}
\begin{flushright}
March 2000
\end{flushright}
\vspace{0.5cm}
\begin{center}
\large
{A Light Sterile Neutrino in the TopFlavor Model}
\end{center}
\begin{center}
{\bf Ehab Malkawi$^a$\footnote{
e-mail:malkawie@just.edu.jo}}, {\bf E.I. Lashin}$^b$, {\bf Hatem Widyan}$^a$\\
$^a${Department of Physics,
Jordan University of Science \& Technology\\
 Irbid 22110, Jordan}\\
$^b${Ain Shams Univ., Cairo, Egypt}
\end{center}
\vspace{0.4cm}
\raggedbottom
\relax
\begin{abstract}
\noindent
A scenario based on the TopFlavor model is presented to explain the origin of a light 
sterile neutrino as indicated by all combined neutrino oscillation experiments. The model 
is phenomenologically well motivated and compatible with all available low-energy data. 
The derived neutrino mass matrix can qualitatively explain
the observed hierarchy in the neutrino mass splittings as indicated by the neutrino 
oscillation data. Numerical results are obtained for special cases. 
\end{abstract}
\vspace{0.5cm}
\vspace{1.0cm}
\end{titlepage}

\section*{I Introduction}

Hints of new physics have recently been advocated through the neutrino sector. 
The recent observation by the 
Super-Kamiokande Collaboration of the atmospheric zenith angle dependent deficit \cite{kam} 
has strengthened  this
conclusion. Also the long term puzzle of the 
solar neutrino deficit \cite{solar} has been a strong demonstration on the existence of 
new physics. 
Recent results on neutrino oscillation have been also reported by 
the Liquid Scintillation Neutrino Detector (LSND) experiment \cite{lsnd}.
The observed anomalies in the neutrino data can be naturally understood in terms of 
massive neutrino oscillations.

The theoretical picture for neutrino oscillation poses a real challenge 
in understanding the form of 
the lepton mass matrix as derived 
from the neutrino oscillation data. A definite picture 
is obscure due to the large number of free parameters in the neutrino and charged 
lepton mass matrices. The phenomenological 
solution is not unique and clearly the need for more data 
and theoretical breakthrough is essential.

Neutrino oscillation can explain both the solar and atmospheric
data in terms of three-generation neutrinos  \cite{bar1}
(ignoring the LSND results \cite{lsnd}.)
In the simplest explanation picture, the solar neutrino data can be understood
in terms of $\nu_e\leftrightarrow \nu_\mu$ oscillation with a small
mass splitting not to influence atmospheric data.
On the other hand, atmospheric data can be explained in
terms of $\nu_\mu\leftrightarrow \nu_\tau$ large mixing with a large
mass splitting compared to the solar case \cite{bar1}.
However, if we combine the LSND result with the solar and atmospheric 
data then we have to include at least
an extra light neutrino. The full oscillation data requires the existence of three different 
scales of 
neutrino mass-squared differences. The different scales can be accommodated only if at least 
a light fourth neutrino exists. 
Such a light neutrino has to be sterile, i.e., to decouple from the low-energy 
observables as indicated by the low-energy experiments \cite{data}.
Some recent phenomenological studies indicate that in the minimal scheme 
the dominant transition of solar neutrinos 
is due to $\nu_e \leftrightarrow \nu_s$ mixing, while the dominant transition of 
atmospheric neutrinos in long-baseline (LBL) experiments is due to 
$\nu_\mu \leftrightarrow \nu_\tau$ mixing \cite{bilenky}.

The inclusion of a sterile neutrino still poses another theoretical challenge. Namely, to understand 
both the origin of the sterile fermion and the very low mass it has. 
The small mass is probably the most
difficult issue in introducing such a particle. 
Any successful scenario has to explain the tiny mass of the
sterile neutrino in a natural way.
A possible scenario would be to generate the light mass through radiative corrections \cite{oka}.
Another interesting scenario would be 
to postulate that the extra neutrino 
is active at a relatively high-energy scale. At that scale the extra neutrino is massless
as the assumed dynamics, due to some symmetry, 
forbids its mass generation. Once the high-energy symmetry is broken (probably in the TeV region), 
a Dirac mass can be generated while the neutrino decouples from the low-energy regime, i.e., becomes
a sterile.
Finally, by invoking the seesaw mechanism we can understand the highly suppressed 
Majorana mass of such a sterile neutrino.

In this work we consider the possibility of understanding the origin of a light sterile neutrino 
through the second scenario. 
A similar model has been discussed in Ref.~\cite{bis}, however, the model suffers from  
theoretical anomalies. Furthermore, an explicit formulation of the model is highly 
complicated. The model we discuss in this work does not suffer from theoretical drawbacks. 
It is based on the gauge nonuniversal 
symmetry $SU(3)_c\times SU(2)_l\times SU(2)_h\times U(1)_Y $
discussed extensively in Refs.~\cite{lima,ehab1}. We refer to this model as the TopFlavor model 
which is anomaly free and phenomenologically well motivated. Several recent 
phenomenological studies have been published in the literature \cite{ehab1}. 
To account for the existence of the sterile neutrino, we modify the standard 
fermion content by the inclusion of few extra fermions. The extra fermion spectrum does not 
appreciably affect the 
low-energy regime because of the heavy mass of the extra active fermions, as discussed later. 
Only one-neutral fermion emerges with 
a small mass while decouples from the low-energy regime which we then call the sterile neutrino.

 The rest of this paper is organized as follows.
 In Sec. II, we briefly review the model. In Sec. III, we enlarge the fermion spectrum
by introducing extra fermions and discuss the mechanism for generating the mass of the 
sterile and active fermions. Finally, in Sec. IV we provide some numerical discussion of the model. 

\section*{II Structure of the Model}

The TopFlavor model \cite{lima,ehab1} is  
based on the gauge symmetry G=~$SU(3)_c\times SU(2)_l\times
SU(2)_h\times U(1)_Y$. In this model, the third generation of fermions 
(top quark $t$, bottom quark $b$, tau lepton $\tau $, and
 its neutrino $\nu_\tau $) is subjected to a new gauge interaction at the high energy scale, 
instead of the usual weak interaction advocated by the standard model (SM) of the electroweak
interaction. On the contrary, the first and second
generations only feel the weak interaction supposedly equivalent to the SM
case. The new gauge dynamics is attributed to the $SU(2)_h$ symmetry
under which the left-handed fermions of the third generation transform in
the fundamental representation (doublets), while they remain to be singlets
under the $SU(2)_l$ symmetry. On the other hand, the left-handed fermions of
the first and second generation transform as doublets under the $SU(2)_l$
group and singlets under the $SU(2)_h$ group. The $U(1)_Y$ group is the SM
hypercharge group. The right-handed fermions only transform under the 
$U(1)_Y $ group as assigned by the SM. Finally, the QCD interactions and the
color symmetry $SU(3)_c$ are the same as that in the SM.

The spontaneous symmetry-breaking of the group 
G=$SU(3)_c\times SU(2)_l \times SU(2)_h\times U(1)_Y$\, is accomplished by introducing the
complex scalar fields $\Sigma$, $\Phi_1$, and $\Phi_2$, where 
$\Sigma\sim (1,2,2,0) $, $ \Phi_1 \sim (1,2,1,1)$, and $\Phi_2\sim (1,1,2,1)$.
For the $\Sigma$ field we explicitly write 
\begin{equation}
\Sigma=\pmatrix{\pi_1^0 & \pi_1^+\cr \pi_2^-& \pi_2^0\cr}\, ,
\label{sigma}
\end{equation}
where all scalar fields are complex.
The group G is then broken in three different stages. 
The first stage of symmetry breaking is accomplished once the $\Sigma $
field acquires a vacuum expectation value (vev) $u$, i.e.,  
$
\left\langle \Sigma \right\rangle =\pmatrix{u & 0 \cr 0 & u \cr}\, ,
$
where $u$ is taken to be real.
The form of $\left\langle \Sigma \right\rangle$  
guarantees the breakdown of $SU(2)_l \times SU(2)_h\rightarrow SU(2)$.
Therefore, the unbroken symmetry  
is essentially the SM gauge symmetry $SU(3)_c\times SU(2)_w\times U(1)_Y$, 
where $SU(2)_w$ is the usual SM weak group. 
At this stage,
three of the gauge bosons acquire a mass of the order $u$, while the other
gauge bosons remain massless. The phenomenology of the model imposes the
constraint $u\gsim 1$ TeV \cite{ehab1} 
The second and third stage of symmetry breaking (the electroweak symmetry-breaking) is  accomplished
through the scalar fields $\Phi_1$ and $\Phi_2$ by
acquiring their vacuum expectation values  
$
\left\langle \Phi_1 \right\rangle=\pmatrix{0 \cr v_1\cr}\,
$,
and $
\left\langle \Phi_2 \right\rangle=\pmatrix{0 \cr v_2\cr}\,
$, respectively. The electroweak symmetry-breaking scale $v$ is defined 
$v\equiv\sqrt{v_1^2+v_2^2}=246$ GeV.
Since the third generation of fermions is heavier than the first two generations, it
is suggestive to conclude that $v_2\gg v_1$.
The surviving symmetry at low energy 
is $SU(3)_c\times U(1)_{\rm{em}}$ gauge symmetry.
Because of this
pattern of symmetry breaking, the gauge couplings $g_l$, $g_h$, and $g_Y$ of $SU(2)_l$, 
$SU(2)_h$, and $U(1)_Y$, respectively, are related to the 
$U(1)_{{\rm {em}}}$ gauge coupling $e$ by the relation 
$
{1/e^2}={1/g_l^2}+{1/g_h^2}+{1/g_Y^2}\,
$ \cite{ehab1}.
We define 
\begin{equation}
g_l=\frac e{\sin \theta \cos \phi }\,,\hspace{1cm} 
g_h=\frac e{\sin \theta \sin \phi }\,,\hspace{1cm}
g_Y=\frac e{\cos \theta }\,,
\end{equation}
where $\theta $ is the usual weak mixing angle and $\phi $ is a new
parameter of the model. 

For $g_h>g_l$ (equivalently $\tan \phi <1$), we require 
$g_h^2\leq 4\pi $ (which implies $\sin ^2\phi \geq g^2/(4\pi )\sim 0.04$) so
that the perturbation theory is valid. Similarly, for $g_h<g_l$, we require 
$\sin ^2\phi \leq 0.96$. For simplicity, we focus on the region where 
$x \equiv u^2/v^2 \gg 1$, and
ignore the corrections which are suppressed by higher powers of $1/x$. 
The light gauge boson masses are found to be 
$
M^2_{W^\pm }=M_Z^2 {\cos ^2{\theta }\,}=
M_0^2(1+O(1/ x) )\,,
$
where $M_0 \equiv ev/2\sin \theta $.
While for the heavy gauge bosons $W^{\prime \pm}$ and $Z^\prime$, one finds \cite{ehab1} 
\begin{equation}
M^2_{{W^\prime}^\pm} = M^2_{Z^\prime }=M_0^2\left( \frac {x}{\sin^2\phi
\cos^2\phi }+O(1) \right) \,.
\end{equation}
Up to this order, the heavy gauge bosons are degenerate in mass 
because they do not mix with the hypercharge gauge boson field.
The SM fermions
acquire their masses through their Yukawa interaction via the 
$\Phi_1 $ and $\Phi_2$ scalar fields. 
For instance, the leptonic Yukawa sector is given by
\begin{eqnarray}
 {\cal L}^\ell_{\mbox{\rm{Yukawa}}} &=&
       \overline{\Psi _L}^1\Phi_1 \left[
       g_{11}^ee_R+g_{12}^e\mu _R+g_{13}^e\tau _R\right] + \nonumber \\
 &&\ \overline{\Psi _L}^2\Phi_1 \left[ g_{21}^ee_R+g_{22}^e\mu _R+g_{23}^e\tau
 _R\right] +\nonumber \\
&&\ \overline{\Psi _L}^3\Phi_2 \left[
 g_{31}^ee_R+g_{32}^e\mu _R+g_{33}^e\tau _R\right] +
 h.c.,  \label{mass_1}
 \end{eqnarray}
where 
\begin{equation}
\Psi_L^1=\pmatrix{\nu_{eL}\cr e_L}\, ,\, 
\Psi_L^2=\pmatrix{\nu_{\mu L}\cr \mu_L}\, , \, \rm{and}\,\, 
\Psi_L^3=\pmatrix{\nu_{\tau L}\cr \tau_L}\, .
\end{equation}

The phenomenology of the model has been studied extensively in 
Refs.~\cite{lima,ehab1} {\footnote{Although some differences in the scalar sector 
exist among those references.}}.
Comparisons with the Large Electron Positron (LEP) and other low-energy data have been investigated and 
constraints on the heavy gauge bosons mass are reported as $M_{W^\prime} \gsim  1$ TeV. 
The parameter $x\equiv u^2/v^2$ is constrained by LEP data to be larger than 20. 
Other low-energy data such as the $\tau$ life time imposes a higher constraint on $x$ for specific 
scenarios of lepton mixing. 
Flavor changing neutral current (FCNC) effects in the lepton and quark sectors have been 
explored and contributions to different processes have been calculated \cite{ehab1}.
Therefore, in this work we only concentrate on the leptonic mass matrix and refer the reader to
Refs.~\cite{lima,ehab1} for a detailed study of the phenomenology of the model. 

\section*{III Extra Fermions}

To explain the existence of the light sterile neutrino in our scenario we 
enlarge the particle spectrum 
by the inclusion of extra fermions. We consider the minimal number of particles 
needed to account for the existence of the light sterile neutrino and without 
introducing anomalies in the structure of the model. Furthermore, consistency with
low-energy data should be maintained and therefore the extra active 
fermions should decouple from the low-energy regime. 
 
At the high-energy scale the gauge symmetry is assumed to be 
G=$SU(3)_c \times SU(2)_l\times SU(2)_h\times U(1)_Y$.
At a lower scale, $\left<\Sigma\right>\sim$ a few TeV, the gauge symmetry 
is broken into the SM symmetry group $H_1=SU(3)_c\times SU(2)_w \times U(1)_Y$.  
At the electroweak scale the final stage of symmetry breaking occurs and the
surviving symmetry group is $H_2=SU(3)_c\times U(1)_{\rm{em}}$. 
The fermion spectrum includes the standard three fermion generations with the 
transformation, under G, as explained in Sec. II. 

We enlarge the fermion spectrum by the inclusion of three sets of extra fermions as follow: 
\begin{itemize}
\item[a)]
To explain the extremely light neutrino masses we invoke the seesaw mechanism. Therefore, 
four right-handed neutrinos are introduced, $\nu_{eR}$, $\nu_{\mu R}$, $\nu_{\tau R}$, and 
$\nu_{s R}$. The four right-handed neutrinos are singlets under G and  
are assumed to be Majorana fermions with masses of the order of the Grand Unified Theory  
(GUT) scale.

\item[b)]
We introduce a bi-doublet fermion field $S_L$ with the transformation
$S_L \sim (1,2,2,0)$. Explicitly, $S_L=\nu_{sL}+S_L^a\tau^a$ transforms, under G, as  
\begin{equation}
S_L\rightarrow g_1 S_L g_2^\dagger \, ,
\end{equation}
where $g_1 \in SU(2)_l$ and $g_2 \in SU(2)_h$. 
Once the symmetry group G is broken down into the symmetry group $H_1$, 
the field $S_L$ decomposes into two parts, with
transformation under $H_1$ as   
$ (1,3,0)+(1,1,0)$.
The neutral field with the transformation $(1,1,0)$ corresponds to the sterile neutrino.  
The triplet field remains an active and thus must acquire a heavy Dirac mass in order
to be consistent with the low-energy data \cite{data}.
To prevent the uncontrolled Majorana mass we assume that $S_L$ carries
a conserved quantum number $z_L$ at the high-energy scale. The new quantum number
is due to a global $U(1)$ symmetry which causes no harmful anomaly to spoil 
the foundation of the model \cite{anomaly}. 

\item[c)]
We introduce
another triplet fermion field $S_R=S_R^a \tau^a$ with the transformation, under G, as $(1,1,3,0)$ 
just for the purpose of giving a Dirac mass to the active triplet field of $S_L$. 
Similar to $S_L$, the field $S_R$ is assumed to carry a
quantum number $z_R$ to prevent the dangerous Majorana mass term. 
Hence, a Dirac mass for the extra active fermions 
is generated through the Yukawa interaction term
\begin{equation}
{\cal L}_a =\frac{g_a}{2}\rm{Tr}\left[ \overline{S_L} \Sigma S_R\right] \, ,
\label{mass1}
\end{equation}
where $g_a\sim 1$ is a Yukawa coupling.
Once the scalar field $\Sigma$ acquires its vev $u\gsim$ 1 TeV, 
a Dirac mass for the triplet fermions is generated, $m_a=g_a u\gsim 1$ TeV.
In order for the Yukawa term in Eq.~(\ref{mass1}) to conserve the assumed
global $U(1)$ symmetry, we require $\Sigma$ to carry a quantum number $z_0$ such 
that $z_L=z_R+z_0$.
\end{itemize}

The sterile neutrino $\nu_{s L}$ acquires its Dirac mass through the Yukawa term
\begin{equation}
{\cal L}_s= \frac{g_s}{2}\rm{Tr}\left[ \overline{S_L}\tau_2 \Sigma^* \tau_2\right] \nu_{sR}\, ,
\label{mass_s}
\end{equation}
where $g_s\sim 1$ is a Yukawa coupling.
The Yukawa terms in Eqs.~(\ref{mass1},\ref{mass_s}) 
conserve the new quantum number provided that we demand 
$z_L=-z_0=z_R/2$.
It is important to notice that the form of $\Sigma$ is  
as given in Eq.~(\ref{sigma}). A simple choice would be 
\begin{equation}
\Sigma=\pi^0 +i\pi^a\tau_a=\pmatrix{\pi^0+i\pi^3 & i\sqrt{2}\pi^+\cr
                                         i\sqrt{2}\pi^- & \pi^0-i\pi^3\cr}\, ,
\end{equation}
where $\pi^0$ and $\pi^a$ are taken to be real fields. However, for this particular 
choice $\tau_2\Sigma^*\tau_2=\Sigma$ 
and therefore, the Yukawa terms in Eqs.~(\ref{mass1},\ref{mass_s}) 
are not simultaneously invariant under 
the assumed global $U(1)$ symmetry.
To conclude, the extra fields we introduce are the minimal number of fields 
required to account for the
existence of a light sterile neutrino without spoiling the accuracy of the 
low-energy data, and also without introducing any theoretical anomalies into the model.

The full neutrino Yukawa interaction terms are given as 
\begin{eqnarray}
 {\cal L}^\nu_{\mbox{\rm{Yukawa}}} &=&
       \overline{\Psi _L}^1{\widetilde{\Phi}}_1 \left[
       g_{11}^\nu \nu_{eR} +g_{12}^\nu \nu_{\mu R}+g_{13}^\nu\nu_{\tau R} + 
       g_{14}^\nu \nu_{sR}\right]+\nonumber \\
 &&\ \overline{\Psi _L}^2 {\widetilde{\Phi}}_1 \left[ g_{21}^\nu \nu_{eR}+g_{22}^\nu \nu_{\mu R} +
    g_{23}^\nu\nu_{\tau R} +g_{24}^\nu \nu_{sR} \right] +\nonumber \\
&&\ \overline{\Psi _L}^3 {\widetilde{\Phi}}_2 \left[
 g_{31}^\nu \nu_{eR}+g_{32}^\nu \nu_{\mu R}+g_{33}^\nu\nu_{\tau R} + 
   g_{34}^\nu \nu_{sR}\right] +\nonumber\\
&& \frac{1}{2}{\rm{Tr}}[\overline{S_L}\tau_2\Sigma^*\tau_2]\left[
 g_{41}^\nu \nu_{eR}+g_{42}^\nu \nu_{\mu R}+g_{43}^\nu\nu_{\tau R} + 
   g_{44}^\nu \nu_{sR}\right] +
 h.c., \label{dirac}
 \end{eqnarray}
where ${\widetilde{\Phi}}_{1,2}\equiv i\tau_2 \Phi_{1,2}^*$.
The Dirac mass matrix derived from Eq.~(\ref{dirac}) is written as
\begin{equation}
M_D=\pmatrix{g_{11}^\nu v_1 & g_{12}^\nu v_1 & g_{13}^\nu v_1 & g_{14}^\nu v_1\cr 
             g_{21}^\nu v_1 & g_{22}^\nu v_1 & g_{23}^\nu v_1 & g_{24}^\nu v_1\cr 
             g_{31}^\nu v_2 & g_{32}^\nu v_2 & g_{33}^\nu v_2 & g_{34}^\nu v_2\cr
             g_{41}^\nu u   & g_{42}^\nu u   & g_{43}^\nu  u  & g_{44}^\nu u\cr} \, .
\label{full}
\end{equation}  
The right-handed neutrino Majorana mass matrix $M_R$ is assumed to have a common 
mass scale of the order of the GUT scale, $M_X\sim 10^{15}$ GeV.
Therefore, the full neutrino mass matrix forms a $8\times 8$ matrix which can be written as
\begin{equation}
M_\nu=\pmatrix{0& M_D\cr M_D^T & M_R\cr}\, .
\end{equation}
By invoking the seesaw mechanism the left-handed neutrino Majorana mass matrix is then given 
as
\begin{equation}
M_L= M_D M_R^{-1} M_D^T\, .
\end{equation} 
Due to the seesaw mechanism all elements of $M_L$ are highly suppressed by the GUT scale $M_X$
of the right-handed Majorana mass matrix $M_R$.  
Therefore, we can introduce the sterile neutrino with a natural mechanism for generating
its light mass as required by the neutrino oscillation data. In the next section we give 
further discussion of the derived neutrino mass matrix.  

\section* {IV Discussion and Conclusions}
The mass matrix in Eq.~(\ref{full}) is written in its most general form. A quantitative analysis 
is attainable only if the structure of the mass matrix
is fully determined which requires further theoretical assumptions to be incorporated in the structure 
of the model.
Nevertheless, the structure of the mass matrix already suggests an interesting behavior.
There are three hierarchical energy scales in the mass matrix $M_D$, namely, $u\gg v_2\gg v_1$
which could be connected to the observed three hierarchical mass scales 
in the neutrino data, namely, 
$\Delta m^2_{\rm{LSND}}\gg \Delta m^2_{\rm{atm.}}\gg \Delta m^2_{\rm{solar}}$.
It is suggestive to conclude that
\begin{eqnarray}
&&\Delta m^2_{\rm{LSND}} \sim {{\left(\frac{u^2}{M_X}\right)}^2}\approx 1 \,{\rm{eV}}^2 \, ,\nonumber \\ 
&&\Delta m^2_{\rm{atm.}} \sim {{\left(\frac{v_2^2}{M_X}\right)}^2}\approx 10^{-3} \,{\rm{eV}}^2\, , 
\nonumber \\
&&\Delta m^2_{\rm{solar}}\sim {{\left(\frac{v_1^2}{M_X}\right)}^2}\approx 10^{-5} \,{\rm{eV}}^2\, ,
\end{eqnarray}
as indicated by the neutrino oscillation data \cite{kam,solar,lsnd,bar1}.
In fact from the LEP data, we already know that $u\gsim v\sqrt{20}\approx 1.2$ TeV \cite{ehab1}.
The observed mass scales can be obtained if we simply choose $u\approx 1.2$ TeV and $v_2\approx 230$ GeV.
From which we conclude that $v_1\approx 75$ GeV, $M_X \approx (1-10)\times 10^{15}$ GeV, and 
\begin{eqnarray}
&&\frac{\Delta m^2_{\rm{LSND}}}{\Delta m^2_{\rm{atm.}}}\sim \frac{u^4}{v_2^4}\approx 10^{+3}\,,\nonumber \\
&&\frac{\Delta m^2_{\rm{LSND}}}{\Delta m^2_{\rm{solar}}}\sim \frac{u^4}{v_1^4}\approx 10^{+5}\,,\nonumber\\
&&\frac{\Delta m^2_{\rm{atm.}}}{\Delta m^2_{\rm{solar}}}\sim \frac{v_2^4}{v_1^4}\approx 10^{+2}\, .
\end{eqnarray}

In the simplest scheme where oscillation data can be explained in terms of two flavor
mixing, it has been argued that the dominant transition of solar neutrino is due to  
$\nu_e \leftrightarrow \nu_s$ mixing \cite{bilenky}. Such a picture can hardly be satisfied
by our model with the above choice of parameters and without the need for fine tuning. 
In the case of $\nu_e \leftrightarrow \nu_s$ mixing one can show that  
the effective $2\times 2$ mass matrix is given as  
\begin{equation}
M_L=\frac{1}{M_X} \pmatrix{g_{1} v^2_1 & g_{2} v_1 u \cr 
             g_{2}v_1 u   & g_{3} u^2   \cr} \, ,
\end{equation} 
where the couplings $g_{1,2,3}\sim 1$. Therefore, one can show that
the solar mass splitting is given as 
$\Delta m^2_{\rm{solar}}\approx u^4/M_X^2$ which is many orders of magnitude larger
than the experimental fit \cite{solar,bar1}.
If we take $M_X$ to be of order $10^{18}$ GeV, which is close to the Planck scale, we get 
a result consistent with the experimental fit as shown in Table 1, where we consider 
the numerical values $u=1200$ GeV, $M_X=10^{18}$ GeV, and $v_1=75$ GeV. 
In Table 1, the numerical values of the Yukawa couplings as well as their solar neutrino solution
are given.
\begin{table}
\begin{center}
\caption{Numerical values for the mass matrix elements and their predicted 
mass splitting and mixing for the case of $\nu_e \leftrightarrow \nu_s$ transition.}
\begin{tabular}{|c|c|c|c|c|}\hline
$g_1$ & $g_2$ & $g_3$ & $\sin^2 2\theta$ & $\Delta m^2$ \\ \hline
0.8& 0.8 & 1.3 & $7\times 10^{-3}$ & $3\times 10^{-6}$\\ 
1.8& 1.0 & 1.5 & $7\times 10^{-3}$ & $4\times 10^{-6}$\\
1.1& 1.0 & 1.6 & $5\times 10^{-3}$ & $5\times 10^{-6}$\\ \hline 
\end{tabular}
\end{center}
\end{table}
However, such a solution is not favored as we can not explain the apparent hierarchy among 
the solar, atmospheric, and LSND data.

In conclusion we have provided a scenario based on the TopFlavor model 
to explain the existence of a light sterile neutrino. 
The scenario is anomaly free and phenomenologically compatible with all existing low-energy data. 
The scenario can also qualitatively explain the hierarchy in the observed mass scales of the 
neutrino oscillation data. 
Quantitative results are obtained for special cases.

\section*{Acknowledgement}
The authors would like to thank G. Senjanovic for useful discussion and comments.
Also they would like to thank ICTP for the kind hospitality where some part 
of this work was done.

\end{document}